\documentstyle[12pt]{article}
\author{M.Arik, T.Rador, \\
Department of Physics,
Bogazici University\\
Bebek 80815,\\
Istanbul Turkey
\and
R.Mir-Kasimov\\
Bogoliubov Laboratory of Theoretical Physics \\
Joint Institute for Nuclear Research\\
Dubna Moscow region                            \\
141980 Russia}
\title{Relativistic SUSY QM as Deformed SUSY QM
}
\begin{document}
\maketitle
\begin{abstract}
{\ The relativistic finite-difference SUSY Quantum Mechanics (QM) is
developed. We show that it is connected in a natural way with the q-deformed
SUSY Quantum Mechanics. Simple examples are given. }
\end{abstract}


 The relativistic QM is based on Snyder's idea of noncommutative
space-time coordinates and the concept of relativistic configurational
space. We refer the reader to the review papers [1-5], and the references
therein. The natural connection of RQM with q-deformations is discussed in
the articles \cite{mir1,mir2}.

The one-dimensional relativistic Schr\"{o}dinger equation has the
form
\begin{equation}
\label{rsch1}
\left( h-e\right) \psi \left( x\right) =
\left(h_{0}+V(x)-e\right) \psi \left( x\right) =0,
\end{equation}
where
\begin{equation}
\label{rham1}
\begin{array}{c}
h_{0}=2mc^{2}\sinh {}^{2}
\frac{i\hbar }{2mc}\frac {d}{dx}=
\frac{\stackrel{\wedge }{k}^{2}}{2m}\rightarrow
-\frac{\hbar ^{2}}{2m}\frac{d^{2}}{dx^{2}}~
in\ non-relativistic\ limit \\  \\
\stackrel{\wedge }{k}\ =
-2mc\sinh {}\frac{i\hbar }{2mc}\frac {d}{dx},\quad
\quad k=2mc\sinh \frac {\chi}{2},
\qquad e=\frac{k^{2}}{2}
\end{array}
\end{equation}
We stress that the relativistic Schr\"{o}dinger equation
$\left( \ref{rsch1}\right) $
is the finite-difference equation with a step equal to the Compton
wave length $\frac \hbar {mc}$ of the particle.
Usually, we shall use the
unit system in which
$\hbar =c=m=1$
It is clear that the factorization method,that plays the key role in SUSY\
QM, must be modified in the finite-difference context \cite{mir1}. Let us
introduce a couple of ladder operators
\begin{equation}
\label{A1}
A^{\pm }=\pm i\sqrt{2}\cdot \alpha \left( x\right) \cdot e^{\pm
\rho \left( x\right) }\sinh \frac {i}{2}\frac {d}{dx}
\cdot e^{\mp \rho \left(x\right) }
\end{equation}
or
\begin{equation}
\label{A2}
A^{\pm }=-i\sqrt{2}\cdot \alpha \left( x\right) \cdot e^{\pm \xi
\left( x\right) }
\left[ \sinh \rho _{\frac {s}{2}}\left( x\right)
\cosh \frac
{i}{2}\frac {d}{dx}\mp \cosh \rho _{\frac {s}{2}}\left( x\right)
\sinh \frac {i}{2}
\frac {d}{dx}\right]
\end{equation}
where
\begin{equation}
\label{rho1}
\rho _{\frac {s}{2}}\left( x\right) =
\sinh \frac {i}{2} \frac {d}{dx}
\rho\left( x\right) \qquad \qquad
\rho _{\frac {c}{2}}\left( x\right) =
\cosh \frac{i}{2}\frac {d}{dx}\rho \left( x\right)
\end{equation}
\begin{equation}
\label{q1}
\xi \left( x\right) =
\rho \left( x\right) -\rho _{\frac {c}{2}}
\left(x\right)
\end{equation}
and $\rho \left( x\right) $ is the logarithm of the ground state wave
function of eq. (\ref{rsch1})
\begin{equation}
\label{rsch2}
\psi _{0}\left( x\right) =e^{-\rho \left( x\right) }
\end{equation}
In the nonrelativistic limit the operators $A^{\pm }$ turn into the
usual ladder operators (cf. \cite{mir1,mir2}).
In addition to the
finite-difference character of the operators (\ref{A1}, \ref{A2}) there are
two factors $\alpha \left( x\right) $ and $e^{\pm \xi \left( x\right) }$
that don't appear in the nonrelativistic case and whose presence enforces
the difference with the nonrelativistic ladder operators.
The factor
$\alpha\left( x\right)$
is connected with a natural lattice variable and is also
expressed in terms of $\rho \left( x\right) $ (see \cite{mir1,sus1} ).
Factors $e^{\pm \xi \left( x\right) }$ are connected with deformations.
Some
quantity (deformation)
\begin{equation}
\label{q2}
q\left( x\right) =e^{a\left( x\right) }
\end{equation}
must be introduced to cancel $e^{\pm \xi \left( x\right) }$.
In the
nonrelativistic case
$\xi \left( x\right) \rightarrow 1$
and there is no
de\-for\-ma\-tion :
$q\left( x\right) \rightarrow 1$.
Let us con\-si\-der the
$q\left(x\right)$-mutator
\begin{equation}
\label{qum1}
\begin{array}{c}
\left[ A^{-},A^{+}\right] _{q\left( x\right) }=
A^{-}\cdot q\left( x\right)
\cdot A^{+}-A^{+}\cdot q^{-1}\left( x\right) \cdot A^{-}= \\
\\
=\frac{\alpha \left( x\right) }{2}
\left\{
\begin{array}{c}
e^{\frac {i}{2}\frac {d}{dx}}\sinh Z\left( x\right)
\alpha \left( x\right)
e^{\frac {i}{2} \frac {d}{dx}}+
e^{-\frac {i}{2} \frac {d}{dx}}\sinh Z\left( x\right)
\alpha \left( x\right) e^{-\frac {i}{2} \frac {d}{dx}}- \\
\\
-e^{\frac {i}{2} \frac {d}{dx}}
\sinh \left( Z\left( x\right) +
2\rho _{\frac {s}{2}}\left( x\right) \right)
\alpha \left( x\right) e^{-\frac {i}{2}\frac {d}{dx}}-
\\
\\
-e^{-\frac {i}{2} \frac {d}{dx}}
\sinh \left( Z\left( x\right) -
2\rho _{\frac {s}{2}}
\left( x\right) \right) \alpha \left( x\right)
e^{\frac {i}{2} \frac {d}{dx}}
\end{array}
\right\}
\end{array}
\end{equation}
where
\begin{equation}
\label{Z1}
Z\left( x\right) =
2\xi \left( x\right) +a\left( x\right)
\end{equation}
Let us recall that the commutator of the nonrelativistic ladder operators
$a^{\pm }$
does not contain the differentiation operators
\begin{equation}
\label{qum2}
\left[ a^{-},a^{+}\right] =
\frac{d^{2}\rho \left( x\right) }{dx^{2}}
\end{equation}
By analogy with
$\left( \ref{qum1}\right) $
we shall require that there is
no finite-difference derivatives
$e^{\pm \frac {i}{2} \frac {d}{dx}}$
in the r.h.s. of (\ref{qum1}).
The simplest way to achieve this is to put
\begin{equation}
\label{Z2}
Z\left( x\right) =0
\end{equation}
The last equation gives the relation connecting
$\rho \left( x\right) $
and
$a\left( x\right) $ (or $q\left( x\right) $):
\begin{equation}
\label{q3}
a\left( x\right) =
-2\xi \left( x\right) \ \qquad q\left( x\right)
=e^{-2\xi \left( x\right) \ }
\end{equation}
We have
\begin{equation}
\label{qum3}
\begin{array}{c}
\left[ A^{-},A^{+}\right] _{q\left( x\right) }=
-2\alpha \left( x\right)
\sinh \frac {i}{2}\frac {d}{dx}
\left[ \alpha \left( x\right) \sinh 2
\rho _{\frac{s}{2}}\left( x\right) \right] \rightarrow  \\
\\
\rightarrow \frac{d^{2}\rho \left( x\right) }{dx^{2}}
\end{array}
\end{equation}
Now let us write down the basic relations of relativistic SUSY QM, i.e.,
the relativistic quantum mechanical system whose Hamiltonian is constructed
of anticommuting charges Q \cite{khare}:
\begin{equation}
\label{suh1}
\stackrel{\wedge }{H}=\frac {1}{2}\cdot
\left\{ Q,Q^{\dagger}\right\} _{q\left( x\right) }=
\frac {1}{2}\cdot \left( Q\cdot q\left( x\right)^{-1}
\cdot Q^{\dagger }+
Q^{\dagger }\cdot q\left( x\right) \cdot Q\right)
\end{equation}
\begin{equation}
\label{charge1}
\left\{ Q,Q\right\} =
\left\{ Q^{\dagger },Q\dagger \right\} =0
\end{equation}
As in the nonrelativistic, case the supersymmetry property of Hamiltonian
$\stackrel{\wedge }{H}$
\begin{equation}
\label{suh2}
\left[\stackrel{\wedge }{H},Q\right] =
\left[\stackrel{\wedge }{H},Q^{\dagger }\right] =0
\end{equation}
is provided by nilpotency of the charge operators
$\left( \ref{charge1}\right) $.
The Hamiltonian $\stackrel{\wedge }{H}$ contains coordinates that
are quantized by $q\left( x\right)$-mutators and anticommutators. They are
mixed by deformed supersymmetry transformations.
The explicit realization of
$Q$ and $Q^{\dagger }$ is
\begin{equation}
\label{charge2}
Q=i\sqrt{2}\cdot A^{+}\cdot
\stackrel{\wedge }{\psi }^{\dagger },\qquad
Q^{\dagger }=
-i\sqrt{2}\cdot A^{-}\cdot \stackrel{\wedge}{\psi }
\end{equation}
In the simplest case the bosonic degrees of freedom represented by the
ladder operators
$A^{\pm }$
are described by the momentum operator (\ref{rham1})
and the position operator $x$ with the commutation relation
\begin{equation}
\label{qum4}
\left[ x,\stackrel{\wedge }{k}\right] =
i\cosh \frac {i}{2} \frac {d}{dx}
\end{equation}
whereas
$\stackrel{\wedge }{\psi }^{\dagger }$
and
$\stackrel{\wedge }{\psi }$
are Fermi degrees of freedom with the corresponding anticommutation
relations:
\begin{equation}
\label{psi1}
\left\{ \stackrel{\wedge }{\psi }^{\dagger },\stackrel
{\wedge }{\psi }\right\} =1,\quad
\left\{ \stackrel{\wedge }{\psi },\stackrel{\wedge }{\psi }\right\} =
\left\{ \stackrel{\wedge }{\psi }^{\dagger },
\stackrel{\wedge }{\psi }^{\dagger }\right\} =0
\end{equation}
This yields (\ref{charge1}) and
\begin{equation}
\label{suh3}
\stackrel{\wedge }{H}=H-\frac {1}{2}\cdot
\left[ \stackrel{\wedge }{\psi }^{\dagger },
\stackrel{\wedge }{\psi }\right] \cdot \Delta V\left(x\right) .
\end{equation}
We introduce the operator
\begin{equation}
\label{suh4}
\begin{array}{c}
\begin{array}{c}
H=\frac {1}{2}\cdot \left\{ A^{-},A^{+}\right\}_{q\left( x\right) }=
\frac{1}{2}\cdot
\left( A^{-}\cdot q\left( x\right) \cdot A^{+}+
A^{+}\cdot q^{-1}\left( x\right) \cdot A^{-}\right) = \\
\\
=H_{0}+
\alpha \left( x\right) \alpha _{\frac {c}{2}}
\left( x\right) -\alpha \left(x\right) \cdot
\cosh \frac {i}{2} \frac {d}{dx} \left( \alpha \left( x\right) \cdot
\cosh 2\rho _{\frac {s}{2}} \left( x\right) \right) \rightarrow  \\
\\
\rightarrow -\frac {1}{2} \frac{d^{2}}{dx^{2}}+
\frac {1}{2} \left( \rho ^{\prime }\left(x\right) \right) ^{2},
\end{array}
\end{array}
\end{equation}
where the operator
\begin{equation}
\label{suh5}
H_{0}=
2\left[ \alpha \left( x\right) \cdot \sinh \frac {i}{2}
\frac{d}{dx}\right] ^{2}=
\frac{\stackrel{\wedge }{P}^{2}}{2}\rightarrow
-\frac {1}{2} \frac{d^{2}}{dx^{2}}
\end{equation}
plays the role of free hamiltonian with the momentum operator
\begin{equation}
\label{rmom1}
\stackrel{\wedge }{P}=
-2\alpha \left( x\right) \cdot
\sinh \frac {i}{2} \frac {d}{dx}
\end{equation}
modified by the interaction
\begin{equation}
\label{suh6}
\begin{array}{c}
H_{+}=A^{+}\cdot q^{-1}\left( x\right) \cdot A^{-}=
H-\Delta V\left( x\right)\\
\\
H_{-}=
A^{-}\cdot q\left( x\right) \cdot A^{+}=
H+\Delta V\left( x\right)
\end{array}
\end{equation}
\begin{equation}
\label{suh7}
\Delta V\left( x\right) =
-\alpha \left( x\right)
\sinh \frac{i}{2}\frac {d}{dx}
\left[ \alpha \left( x\right) \sinh 2\rho _{\frac {s}{2}}
\left(x\right) \right]
\end{equation}
In the
$\left( 2\times 2\right)$-representation
\begin{equation}
\label{psi2}
\stackrel{\wedge }{\psi }^{\dagger }=
\sigma _{-}=
\left(\begin{array}{cc}
0 & 0 \\
1 & 0
\end{array}
\right) ,\qquad \stackrel{\wedge }{\psi }=
\sigma _{+}=\left(
\begin{array}{cc}
0 & 1 \\
0 & 0
\end{array}
\right)
\end{equation}
\begin{equation}
\label{psi3}
\left[ \stackrel{\wedge }{\psi }^{\dagger },
\stackrel{\wedge }{\psi }\right] =
-\sigma _{3}
\end{equation}
we find from (\ref{suh3})
\begin{equation}
\label{suh8}
\begin{array}{c}
\stackrel{\wedge }{H}=
H+\frac {1}{2}\cdot
\Delta V\left( x\right) \cdot \sigma_{3}= \\  \\
=\left(
\begin{array}{cc}
H_{-} & 0 \\
0 & H_{+}
\end{array}
\right) =\left(
\begin{array}{cc}
A^{-}\cdot q\left( x\right) \cdot A^{+} & 0 \\
0 & A^{+}\cdot q^{-1}\left( x\right) \cdot A^{-}
\end{array}
\right)
\end{array}
\end{equation}
 {\bf EXAMPLES}          \\
 {\em 1) RELATIVISTIC OSCILLATOR} (q-oscillator)
\cite{mir1}- \cite{arik}.\\ In this case, we have \cite{mir1,mir2}:
\begin{equation}
\label{rho11}
\rho \left( x\right) =
\frac{m\omega x^{2}}{2\hbar }
\end{equation}
The deformation parameter is a constant and we come to the
$q$-oscillator
with
\begin{equation}
\label{qu1}
q\left( x\right) =
const=q=
e^{-\frac{\omega \hbar }{4mc^{2}}}
\end{equation}
and
\begin{equation}
\label{alpha1}
\alpha \left( x\right) =
\frac {1}{\cos \frac{\omega x}{2c}}.
\end{equation}
The finite-difference ladder operators have the form
\begin{equation}
\label{A3}
A^{\pm }=
\pm i\sqrt{2}\cdot e^{\pm \frac {\omega}{8}}\cdot
\left(\sinh \frac {i}{2}\frac {d}{dx}
\mp i\tan \frac{\omega x}{2}\cdot \cosh \frac {i}{2}
\frac{d}{dx}\right) .
\end{equation}
SUSY Hamiltonian (\ref{suh8}) becomes
\begin{equation}
\label{suh9}
\begin{array}{c}
\stackrel{\wedge }{H}=
\left(
\begin{array}{cc}
e^{-\frac {\omega}{4}}A^{-}A^{+} & 0 \\
0 & e^{\frac {\omega}{4}}A^{+}A^{-}
\end{array}
\right) =\left(
\begin{array}{cc}
h+e_{0} & 0 \\
0 & h-e_{0}
\end{array}
\right) \rightarrow  \\
\\
\rightarrow \left(
\begin{array}{cc}
-\frac {1}{2}
\frac{d^{2}}{dx^{2}}+
\frac{\omega ^{2}x^{2}}{2}+
\frac {\omega}{2} & 0 \\
0 & -\frac {1}{2}\frac{d^{2}}{dx^{2}}+
\frac{\omega ^{2}x^{2}}{2}-
\frac {\omega}{2}
\end{array}
\right)
\end{array}
\end{equation}
where
\begin{equation}
\label{e0}
e_{0}=
2\sinh \frac {\omega}{4}\rightarrow
\frac {\omega}{2}
\end{equation}
and
\begin{equation}
\label{suh10}
\begin{array}{c}
h=\left\{ A^{-},A^{+}\right\} _{q}=
2\left\{ \left( \frac {1}{\cos\frac{\omega x}{2}}
\cdot \cosh \frac {i}{2}\frac {d}{dx}\right) ^{2}-
\cosh \frac {\omega}{4}\right\} = \\  \\
=\frac{\stackrel{\wedge }{P}}{2}+V\left( x\right)
\end{array}
\end{equation}
\begin{equation}
\label{rmom2}
\stackrel{\wedge }{P}=
-\frac {2}{\cos \frac{\omega x}{2}}\cdot
\sinh \frac {i}{2}\frac {d}{dx}\rightarrow -
i\frac {d}{dx}
\end{equation}
The relativistic oscillator potential is
\begin{equation}
\label{Vosc}
V\left( x\right) =
\frac{\cosh \frac {\omega}{4}\cdot
\left[ \sin{}^{2} \frac{\omega x}{2}-
\sinh {}^{2}\frac{\omega}{4}\right]}
{\left[ \cos {}^{2}\frac{\omega x}{2}+
\sinh {}^{2}\frac {\omega}{4}\right] }\rightarrow
\frac{\omega ^{2}x^{2}}{2}
\end{equation}
The spectrum is
\begin{equation}
\label{spectr1}
\begin{array}{c}
\left(
\begin{array}{cc}
e_{n}^{-}=
2\left( e^{\frac{2n+1}{4}\omega }-
e^{-\frac {\omega}{4}}\right)  & 0 \\
0 & e_{n}^{+}=
2\left( e^{\frac{2n+1}{4}\omega }-
e^{\frac {\omega}{4}}\right)
\end{array}
\right) \rightarrow  \\
\\
\rightarrow \left(
\begin{array}{cc}
\left( n+\frac {1}{2}\right) \omega +\frac {\omega}{2} & 0 \\
0 & \left( n+\frac {1}{2}\right) \omega -\frac {\omega}{2}
\end{array}
\right)
\end{array}
\end{equation}
Thus, we have two $q$-oscillators with zero point of energy shifted by
$\pm e_{0}$.
In other words, the energies of
$q$-supersymmetric partners are
connected by
\begin{equation}
\label
{spectr2}
e_{n+1}^{-}=
q^{-2}\cdot e_{n}^{+}
\end{equation}
 {\em 2) THE RADIAL PART OF THREE - DIMENSIONAL RELATIVISTIC\\
SCHR\"{O}DINGER EQUATION}
\cite{kad2,kad3,amir,mir1}:\\
This equation
\begin{equation}
\begin{array}{c}
\label{hrad1}
H_{l}s_{l}\left( r,\chi \right) =
\left\{ 2\sinh {}^{2}\frac {i}{2}
\frac{d}{dr}+
\frac{l\left( l+1\right) }{2r\left( r+i\right) }
e^{i\frac{d}{dr}}\right\} s_{l}\left( r,\chi \right) = \\
=\left( \cosh \chi -1\right)
s_{l}\left( r,\chi \right)
\end{array}
\end{equation}
can be considered as one-dimensional with the potential
$\frac{l\left(l+1\right) }{2r\left( r+i\right) }
e^{i\frac {d}{dr}}$.
Solutions of this
equation, i.e., the free relativistic radial waves have the form
\begin{equation}
\label{fsol1}
s_{l}\left( r,\chi \right) =
\sqrt{\frac{\pi \sinh \chi }{2}}\cdot
\left( -i\right) ^{l+1}\cdot
\frac{\Gamma \left( ir+l+1\right) }
{\Gamma\left( ir\right) }\cdot P_{ir-\frac {1}{2}}^
{-\left( l+\frac {1}{2}\right) }
\left(\cosh \chi \right)
\end{equation}
In the nonrelativistic limit, these functions turn into free solutions of
the Schr\"{o}dinger equation
\begin{equation}
\label{fsol2}
s_{l}\left( r,\chi \right) \rightarrow
s_{l}\left( pr\right) =
\sqrt{\frac{\pi rp}{2}}\cdot
J_{l+\frac {1}{2}}\left( pr\right)
\end{equation}
In this case the relativistic finite-difference ladder operators have the
form
\begin{equation}
\label{A4}
\lambda ^{\pm }=
\pm \frac {i}{\sqrt{2}}
\left[ \frac{ir\mp \left(l+1\right) }{ir}-
e^{i\frac {d}{dr}}\right] \rightarrow a^{\pm }=
\mp \left[\frac {d}{dr}\pm \frac{l+1}{r}\right]
\end{equation}
Hence,
\begin{equation}
\label{suh11}
H_{+}=H_{l}=
\lambda ^{+}\cdot e^{i\frac {d}{dr}}\cdot
\lambda^{-}\qquad H_{-}=
H_{l+1}=\lambda ^{-}\cdot
e^{i\frac {d}{dr}}\cdot \lambda^{+}
\end{equation}
In contrast with the nonrelativistic case, the rising and lowering operators
$\Lambda ^{\pm}$,
which shift the value of the angular momentum
\begin{equation}
\label{A5}
\Lambda ^{+}s_{l+1}\left( r,\chi \right) =
s_{l}\left( r,\chi \right)
\qquad \Lambda ^{-}s_{l}\left( r,\chi \right) =
s_{l+1}\left( r,\chi \right)
\end{equation}
and the ladder operators (\ref{A4}) factorizing the Hamiltonian are
different:
\begin{equation}
\label{A6}
\begin{array}{c}
\Lambda ^{+}= \frac {i}{\sinh \chi }\cdot
\left[ \cosh \chi -\frac{ir-l-2}{ir-1}\cdot
e^{i\frac {d}{dr}}\right] \rightarrow a^{+} \\  \\
\Lambda ^{-}=-\frac i{\sinh \chi }\cdot
\left[ \cosh \chi -\frac{ir+l}{ir-1}
\cdot e^{i\frac {d}{dr}}\right] \rightarrow a^{-}
\end{array}
\end{equation}
Let us consider the identity
\begin{equation}
\label{ident1}
\left[ -H_{l+1}+H_{l}\right] \cdot
H_{l}-\left[ H_{l+1}-H_{l}\right]
\cdot H_{l}\equiv 0
\end{equation}
Using the relation
\begin{equation}
\label{ident2}
\lambda ^{-}e^{i\frac {d}{dr}}-
e^{i\frac {d}{dr}}\lambda^{-}=
-\frac {i}{\sqrt{2}}\cdot \left[ H_{l+1}-H_{l}\right]
\end{equation}
we have
\begin{equation}
\label{ident3}
\left[ -H_{l+1}+H_{l}\right] \cdot
H_{l}-i\sqrt{2}\cdot
\left[\lambda ^{-}e^{i\frac {d}{dr}}-
e^{i\frac {d}{dr}}\lambda ^{-}\right] \cdot H_{l}=0
\end{equation}
After acting on
$s_{l}\left( r,\chi \right) $
and taking into account (\ref{hrad1}) and the relation
\begin{equation}
\label{ident4}
\Lambda ^{-}=
\frac {i}{\sinh \chi }\cdot
\left[ -\left( \cosh\chi -1\right) +
i\sqrt{2}\cdot e^{i\frac {d}{dr}}\cdot
\lambda ^{-}\right],
\end{equation}
we come to the formula
\begin{equation}
\label{ident5}
H_{l+1}\cdot \Lambda ^{-}s_{l}
\left( r,\chi \right) =
\Lambda^{-}\cdot H_{l}s_{l}\left( r,\chi \right)
\end{equation}
which allows us to consider relativistic
$l$ and $l+1$
states as deformed
supersymmet\-ric partner states.
If $s_{l}\left( r,\chi \right) $
is the eigenstate of $H_{l}$, then
$\left( \Lambda ^{-}s_{l}\left( r,\chi \right)\right) $
is the eigenstate of $H_{l+1}$ with the same eigenvalue
\begin{equation}
\label{ident6}
\begin{array}{c}
H_{l}s_{l}\left( r,\chi \right) =
\left( \cosh \chi -1\right) \cdot s_{l}
\left(r,\chi \right) \rightarrow \\
\\
\rightarrow H_{l+1}\cdot \left( \Lambda ^{-}s_{l}
\left( r,\chi \right) \right)=
\left( \cosh \chi -1\right) \cdot
\left( \Lambda ^{-}s_l\left( r,\chi\right) \right)
\end{array}
\end{equation}
The nonrelativistic (undeformed ) analog of (\ref{ident5}) is the relation
\begin{equation}
\label{ident7}
H_{l+1}\cdot a^{-}s_{l}\left( pr\right) =
a^{-}\cdot H_{l}s_{l}\left(pr\right)
\end{equation}


\begin{thebibliography}{99}
\bibitem{kad1}  V.G.Kadyshevsky, in book {\em Problems of
theoretical Physics}, dedicated to the memory of I.E.Tamm {\em Nauka
Publishers}, Moscow (1972) 52. \\ \vspace*{-0.7cm}

\bibitem{kad2}  V.G.Kadyshevsky, R.M.Mir-Kasimov and N.B.Skachkov,
{\em Nuovo Cimento} {\bf 55 A} (1968) 233. \\ \vspace*{-0.7cm}

\bibitem{kad3}  V.G.Kadyshevsky, R.M.Mir-Kasimov and N.B.Skachkov,
{\em Physics of Elementary Particles and Atomic Nucleus} {\bf 2}, N3, (1972)
635 \\ \vspace*{-0.7cm}

\bibitem{amir}   I.V.Amirkhanov, G.V.Grusha and R.M.Mir-Kasimov, {\em %
Physics of Elementary Particles and Atomic Nucleus} {\bf 12}, N3, (1981) 651
\\ \vspace*{-0.7cm}

\bibitem{mir1}   R.M.Mir-Kasimov, {\bf Preprint} SISSA 197/94/EP,
(1994) \\ \vspace*{-0.7cm}

\bibitem{mir2}  R.M.Mir-Kasimov, {\bf Preprint}
Centre de Recherches
Mathematiques,
Uni\-ver\-si\-te de Mont\-re\-al, CRM-2186, (1994);
E.D.Kagramanov,
R.M.Mir-Kasimov and Sh.M.Nagiyev, {\em J.Math.Phys.}, {\bf 31}, (1990) 1733;
R.M.Mir-Kasimov, {\em J. Phys. A} {\bf 24} (1991) 4283 \\ \vspace*{-0.7cm}
\bibitem{macf}  A.J.Macfarlane, {\em J.Phys.A}{\bf 22} (1989) 4581;
L.C.Biedenharn ibid {\bf 22} (1989) L873. \\ \vspace*{-0.7cm}

\bibitem{arik}  M.Arik, Ph.D. Thesis, University of Pittsburgh
(1974) unpublished; M.Arik, D.D.Coon, J.Math.Phys. 17 (1975) 524; M.Arik,
M.Mungan, Phys.Lett.B {\bf 282} (1992) 101. \\ \vspace*{-0.7cm}

\bibitem{sus1}  A.F.Nikiforov, S.K.Suslov, V.B.Uvarov {\em Classical
Orthogonal Polynomials of a Discrete Variable} Springer-Verlag,
Berlin-Heidelberg, (1991); G.Gasper and M.Rahman {\em Basic Hypergeometric
Series}, Cambridge University Press (1990) \\ \vspace*{-0.7cm}

\bibitem{khare}  F.Cooper, A.Khare, U.Sukhatme, {\em Phys.Rep.}{\bf
251}, N 5 \& 6, (1995); F.Schwabl {\em Quantum Mechanics}, Springer-Verlag
(1992) \\ \vspace*{-0.7cm}

\bibitem{fil}  A.T.Fi\-lip\-pov, A.P.Isa\-ev, {\em Mod.Phys.Lett.},{\bf A4}
, (1989), 2167; A.P.Isaev and
R.P.Malik, {\em Phys.Lett.B}{\bf 280}, (1992), 219; V.I.Manko,
G.Marmo, S.Solimeno and F.Zaccaria, {\em Phys.Lett.A} {\bf 176},(1993) 173;
L.A.Slepchenko,{\em Theor. and Math.Phys.}{\bf 78}(1989) 211 ; V.Spiridonov,
in the Proc. of Workshop on Harmonic Oscillators (College Park, USA, 25-28
March 1992). Eds. D.Han, Y.-S.Kim, and W.W.Zachary, NASA Conf. Publ. 3197,
199, pp. 93-108; A.N.Sissakian, V.M.Ter-Antonyan, G.S.Pogosyan,
I.V.Lutsenko, {\em Phys.Lett A}{\bf 143} (1990) 247; M.Chaichian, P.P.Kulish
and J.Lukierski, {\em Phys.Lett} {\bf B237} (1990) 401;
\end{thebibliography}
\end{document}